\documentclass[11pt,preprint]{article}
\usepackage{graphicx}
\usepackage[dvipdfm,margin=1.5in,inner=1in,outer=0.7in,headheight=14pt,bottom=1in,footskip=0.3in,headsep=0.1in,letterpaper]{geometry}
\usepackage{mathrsfs}
\usepackage{url}
\usepackage{multirow}
\usepackage{array}

\title{Limits in late time conversion of cold dark matter into dark radiation}

\author{D. Boriero, P. C. de Holanda and M. Motta\\ Instituto de F\'isica Gleb Wataghin -- UNICAMP, 13083-859, Campinas SP, Brazil \\ danielb@ifi.unicamp.br, holanda@ifi.unicamp.br, mmotta@ifi.unicamp.br}

\begin{document}

\maketitle

\begin{abstract}
Structure formation creates high temperature and density regions in the Universe that allow the conversion of matter into more stable states, with a corresponding emission of relativistic matter and radiation. An example of such a mechanism is the supernova event, that releases relativistic neutrinos corresponding to $99\%$ of the binding energy of remnant neutron star. We take this phenomena as a starting point for an assumption that similar processes could occur in the dark sector, where structure formation would generate a late time conversion of cold dark matter into a relativistic form of dark matter. We performed a phenomenological study about the limits of this conversion, where we assumed a transition profile that is a generalized version of the neutrino production in supernovae events. With this assumption, we obtained an interesting modification for the constraint over the cold dark matter density. We show that when comparing with the standard $\Lambda$CDM cosmology, there is no preference for conversion, although the best fit is within $1\sigma$ from the standard model best fit. The methodology and the results obtained qualify this conversion hypothesis, from the large scale structure point of view, as a viable and interesting model to be tested in the future with small scale data, and mitigate discrepancies between observations at this scale and the pure cold dark matter model.
\end{abstract}

\section{\label{sec:introduction}Introduction}

Cosmological observations accumulated over the last decades have been favoring a picture of the Universe with $25\%$ of the energy density made up of a weakly interacting non-baryonic particle. A scenario where cold dark matter (CDM) dominates is consistent with a bottom-up structure formation, therefore is favored by observations. This comes from the fact that the size of the initial perturbations is limited by the free-streaming length of the dark matter (DM) particles, for it is not possible to have gravitational clustering on scales larger then this length.

The $\Lambda$CDM model with cold dark matter and a cosmological constant still provides the most simple description for the observations that nicely fits the data, e.g. the anisotropies of the Cosmic Microwave Background (CMB) measured by the Wilkinson Microwave Anisotropy Probe-7yr (WMAP)~\cite{0067-0049-192-2-18}, the matter power spectrum tracked by Luminous Red Galaxies (LRG) and measured by Sloan Digital Sky Survey-DR8 (SDSS)~\cite{MNR:MNR16276} and the recent accelerated expansion detected by the standard candles, supernova type Ia (SNIa), and collected by Union-2 compilation (Union-2)~\cite{0067-0049-185-1-32}. But, in spite of the $\Lambda$CDM success, there are still reasons to be cautious. A cosmological constant suffers from a serious fine-tuning and a coincidence problem~\cite{PhysRevLett.59.2607,PhysRevLett.74.846,PhysRevLett.82.896} and, besides the dark energy puzzle, we haven't yet directly detected a dark matter particle, if there is any. Furthermore, specially regarding the small scales, many controversies remain unsettled, e.g. the core-cusp~\cite{maccio,springel} and the missing satellites problem~\cite{Kazantzidis:2011xi,Wang:2012bt}. Generally, simulations of structures produce dark matter halos with a very steep profile, not matching observations, and the number of substructures around galaxies predicted by simulations is superior than the observed one. Efforts have been made in an attempt to improve both simulations and observations, but a general disagreement remains between approaches and results from different groups. 

Naturally, many alternative scenarios have been proposed to deal with the small scales issues, where dark matter, or at least a fraction of dark matter is not cold. The free streaming length of the DM particle limits the size of structures, setting the mass of a warmer candidate around the keV scale in a scenario in which this candidate is dominant. A dark matter particle with this mass scale could be a sterile neutrino~\cite{0004-637X-714-1-652}. Some recent proposals of warm and warm$+$cold dark matter cosmologies have shown that a standard warm dark matter (WDM) cosmology does not solve the core-cusp problem~\cite{Maccio:2012qf} and, while WDM would naturally lead to a smaller number of satellites~\cite{Lovell:2011rd}, other scenarios are not excluded, like the decay of cold thermal relics~\cite{Doroshkevich:1989bf,Strigari:2006jf}. 
On the other hand, the effect of baryons potentially plays a significant role in the galaxy dynamics~\cite{TrujilloGomez:2010yh} and still presents a challenge for simulations.  Accounting for these effects can lower the number of subhalos as 
pointed out in ref.~(\cite{Mashchenko11012008,Governato:2009bg}), however it is not yet certain whether a CDM$+$baryonic processes can reconcile simulations and observations of Milky Way satellites. More recently, there are claims~\cite{Governato:2012fa} that gas outflows from stellar activity can significantly attenuate the steepness of the core. A proposal~\cite{PhysRevLett.109.231301} of long range interactions between cold dark matter particles also would be able to dynamically settle the issue. Still in the spirit of mixed scenarios, it is interesting to see how cosmology can constrain limits on hot dark matter particles, like neutrinos or axions~\cite{Hannestad:2010yi}.

In this work we do not assess the particle nature of dark matter. We suppose that a cosmological scenario allows for a certain amount of dark radiation and consider that this relativistic component is the result of a general conversion process. Naturally, the particles are not collisionless, but we don't make any restrictions about the interactions that could lead to conversion. Our work is inspired on processes in the baryonic sector that convert non-relativistic into relativistic matter. Known astrophysical phenomena can cause such conversions, like accretion disks around black holes and supernovae events, the latter one being the most interesting for us due to the large amount of energy released as relativistic neutrinos. 

When the gravitational pressure of a star exceeds the degeneracy pressure of electrons, the star explodes (or collapses) releasing radiation and neutrinos. The energy budget of the relativistic species is mostly carried by the neutrinos. Typically, this energy corresponds to $99\%$ of the binding energy of the remnant neutron star. 
We take the net effect of this phenomena as a starting point for a supposition that similar processes can occur in the dark sector, and perform a phenomenological study about the limits in a late time and environmental conversion of cold dark matter into dark radiation. There are many different models with such a transition, e.g. in ref.~(\cite{PhysRevD.83.063520,PhysRevD.86.083529}), where dark matter decays at earlier times and in ref.~(\cite{Amendola:2007yx,Ayaita:2012xm}) where a coupling neutrino-quintessence accelerates neutrinos to relativistic velocities in scales of order of Mpc in the recent Universe.

This paper is organized as follows. In section \ref{sec:conversionmodel} we describe the model building. We present a model for the integrated diffuse neutrino flux in the Universe. Afterwards, we generalize the model for the conversion of cold dark matter to dark radiation introducing the two parameters that describe the model: the initial time of the conversion $a_i$ and the conversion rate $\eta$. 
In section \ref{sec:equations} we derive the Boltzmann equations that governs the dark matter particle's distribution, obtaining an equation for the energy density of both components.
In section \ref{sec:analysis} we describe the statistical analysis using Markov Chain Monte Carlo (MCMC) with the up to date data on large scale structure. Finally, in section \ref{sec:results} and \ref{sec:conclusions} we discuss our results and prospects.
 
\section{\label{sec:conversionmodel}Supernova production rate}

Most models of supernova neutrino flux are built from direct observation of core collapse supernovae. However this method only provides a good parametrization for small redshifts, and since we are interested in the effects of such production rate in all cosmological history, we will use a model where the time-dependent supernova rate $R_{SN}(z)$ is proportional to the star formation rate $R_{SF}(z)$~\cite{0004-637X-498-1-106,0004-637X-548-2-522,Ando:2004hc,Ando:2005ka,Lunardini:2005jf}
\begin{equation}
R_{SN}(z)=\frac{\int_{8M_\odot}^{50M\odot}\phi(m)dm}
{\int_{0.1\odot}^{125M_\odot}m\phi(m)dm}R_{SF}(z) \ ,
\end{equation}
and the star formation rate is parametrized as
\begin{equation}
R_{SF}(z)=h\,0.3 \frac{e^{3.4z}}{e^{3.8z}+45} [M_\odot .\textrm{yr}^{-1} \textrm{Mpc}^{-3}] \ .
\end{equation}
We normalize the rate by considering the number of stars that would most likely undergo a supernova event with a correspondent emission of neutrinos, from $8M_\odot$ to $50M\odot$. This can be obtained through the integration of the initial mass function 
\begin{equation}
\phi(m)\propto
\left\{
  \begin{array}{lcl}
    m^{-2.35} & ; & 1 M_\odot < m \\
    m^{-2.33-1.82\log{m}} & ; & 0.1 M_\odot < m < 1 M_\odot
  \end{array} 
\right. \ .
\end{equation}
where we used the combined Salpeter~\cite{salpeter} and Gould~\cite{gould} initial mass functions in the appropriate mass intervals~\cite{madauIMF}.

\begin{figure}\begin{center}        
\includegraphics[width=0.7\linewidth]{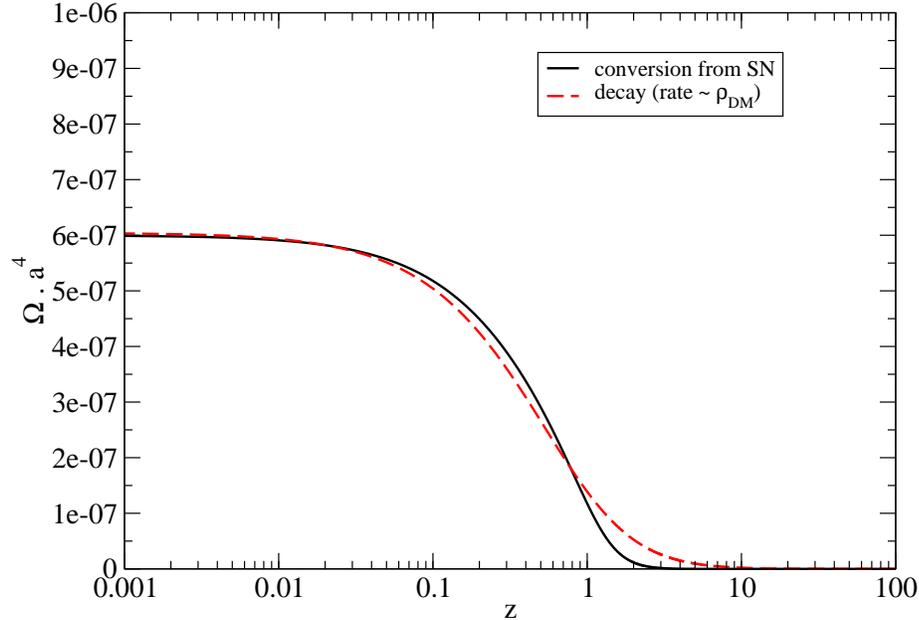}
\caption{The supernova produced neutrino density fraction, $\Omega=\rho_h/\rho_{cr}$, as a function of $z$.}
\label{fig:nudensity}
\end{center}\end{figure}

As it is well known, relativistic neutrinos are the main energetic output from a supernova explosion, where $\sim 99\%$ of the progenitor gravitational binding energy is transferred to neutrinos. As a consequence, a diffuse supernova neutrino flux is present in our Universe. The energy density carried by these neutrinos can be calculated from the evolution equations:
\begin{eqnarray}
\dot{\rho}_b&=&-3H\rho_b - \bar{E}_\nu R_{SN} \ , \\
\dot{\rho}_\nu&=&-4H\rho_\nu+\bar{E}_\nu R_{SN} \ ,
\end{eqnarray}
where $H=\dot{a}/a$ is the Hubble parameter and $\bar{E}_\nu\sim 3\times 10^{53}$ erg is the total energy carried by the neutrino flux in a supernova explosion.  The solutions to these equations are given by:
\begin{eqnarray}
\rho_b&=&\frac{1}{a^3}\left[\rho_0-\int^aa^3\bar{E}_\nu R_{SN}dt \right] \ , \\
\rho_\nu&=&\frac{1}{a^4}\int^aa^4\bar{E}_\nu R_{SN}dt \ .
\end{eqnarray}
In figure \ref{fig:nudensity} we can see the profile of energy density carried by relativistic neutrinos produced by SN explosion as a function of redshift. In the same plot we show the energy density carried by  relativistic species created in dark-matter decay, where the parameters were chosen to produce the same amount of energy density of relativistic species. It is possible to see from figure \ref{fig:nudensity} that two very different mechanisms can produce similar patterns of relativistic species production. 
We propose a scenario where cold dark matter is converted into dark radiation with a not initially populated distribution by a generic mechanism that is not necessarily the decay of the original cold dark matter particle content, but resembles the production mechanism of supernova neutrino. We chose the following \textit{ansatz} for the cold dark matter density evolution 
\begin{equation}
\rho_ca^3=\left\{\begin{array}{ll}\rho_0 & ;a<a_i \\ \rho_0 e^{-\eta(a-a_i)} & ;a>a_i\end{array}\right.  \ , \label{eq:cold}
\end{equation}
where $a_i$ defines a starting time for the conversion and $\eta$, the rate that this conversion occurs. The evolution equations can then be easily integrated, and we obtain for the relativistic dark matter density
\begin{equation}
\rho_ha^4=\left\{\begin{array}{ll}0 & ;a<a_i \\ \rho_{c_0}\left(a_i+\frac{1}{\eta}\right)\times\left[1 \ - e^{-\eta(a-a_i)}\right] & ;a>a_i \end{array} \right. \ . \label{eq:hot}
\end{equation}
With this parametrization the conversion profile on SN neutrinos can be fairly reproduced by the the choice $a_i=0.3$ and $\eta=4.6\times 10^{-6}$, indicating that our parametrization presents a good versatility and a close connection to known astrophysical processes where a conversion from non-relativistic to relativistic matter occurs.

\begin{figure}\begin{center}
\includegraphics[width=0.7\linewidth]{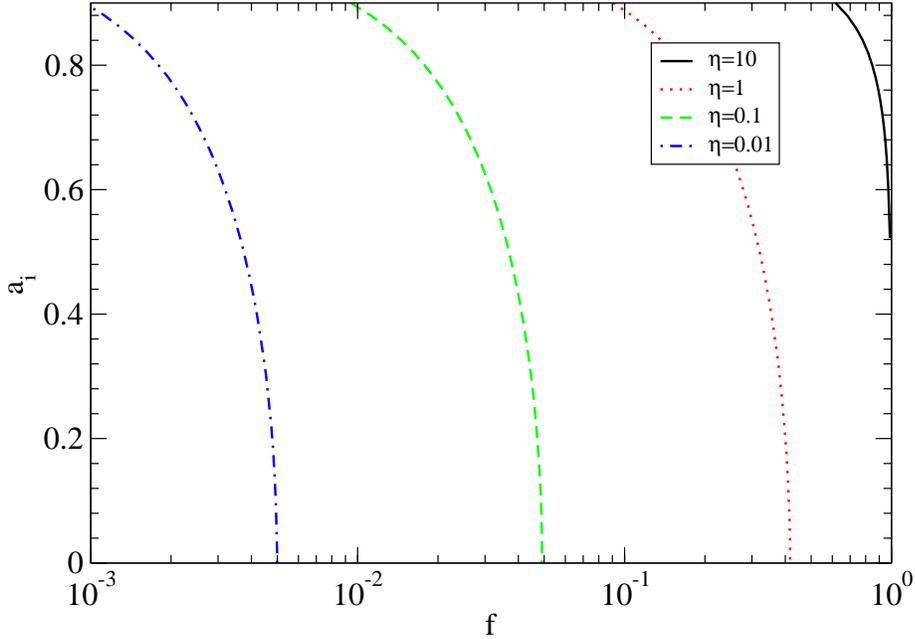}
\caption{Values of parameter $\eta$ from $f$ and $a_i$.}
\label{fig:etafromf}
\end{center}\end{figure}

When constraining our parameters, it is more convenient to work with variables that gives us directly the amount of dark matter in a relativistic form today. This can be calculated explicitly by:
\begin{equation}
f=\left.\frac{\rho_h}{\rho_c+\rho_h}\right|_{a=1}~~~.
\end{equation}
The correspondence between the parameter $\eta$ and $f$, for a given value of $a_i$, is unique, and can be obtained numerically from the relations above. We present this correspondence in figure \ref{fig:etafromf}.

\section{Boltzmann equations}\label{sec:equations}
Perturbations over the homogeneous background of the cosmic ingredients can be evolved as a series of coupled Boltzmann equations of perfect fluids, where each component is fully described as a hierarchized set of equations, truncated according to its thermal condition and coupled according to the physics considered. For a complete description of all the components, see the ref.~(\cite{Ma:1995ey}). In this work, it is going to be detailed only the set of equations of cold dark matter and dark radiation with the environmental conversion as a coupling. We are supposing that the cold dark matter is a particle with mass $m_c$ that undergoes an environmental conversion at a rate $dg_c/da$ to a dark radiation with an empty distribution populated at a rate $dg_h/da$. The Boltzmann equation for the cold dark matter is
\begin{equation}
 \frac{d f_c}{d t}  =C\left[ f_c\right] .
\end{equation}
In this scenario with dynamical dark matter, the total distribution function depends on the momentum and the scale factor, we also consider that the dependencies are detachable
\begin{equation}
 C[f_c(p,a)] =-\dot{a}f_c(p)\frac{dg_c(a)}{da} ,
\end{equation}
where the function $g_c(a)$ represents an effective and already normalized conversion process. We consider that the phase space of the daughter particle is completely available. Integrating the distribution function, the final equations of density will be
\begin{equation}
 \dot{{\rho}}_c + 3\frac{\dot{a}}{a}{\rho}_c = -a\dot{a}\frac{dg_c(a)}{da}{\rho}_c , \label{eq:cdm}
\end{equation}
and density perturbations are
\begin{equation}
 \dot{{\delta}}_c + ik{v_c} + 3\dot{{\Phi}} =-a\dot{a}\frac{dg_c(a)}{da} {\delta}_c. 
\end{equation}
For non-relativistic particles is enough to extend the hierarchical description to the velocity term 
\begin{equation}
 \dot{{v_c}} + \frac{\dot{a}}{a} {v_c} + ik {\Psi} =-a\dot{a}\frac{dg_c(a)}{da}{v_c} ,
\end{equation}
any higher order as pressure or stress is negligible and even undesirable, since there is no seen pressure effects in the clustering of dark matter. In a similar way, we detail the equations of density and density perturbations of the dark radiation, which is identical to the massless neutrino equations except for a non-null collision term
\begin{equation}
 \frac{d f_h}{d \tau}  =C\left[ f_h\right] .
\end{equation}
The distribution of relativistic particles in thermodynamical equilibrium is completely described by the temperature. We assume that this is the case and include the perturbations directly in the distribution function
\begin{equation}
 f_h(x^i,p_j,\tau) ={f_h}_0(p,\tau)\times[1 + \Psi(x^i,p,n_j,\tau)] . \
\end{equation}
The collision term populates the dark radiation from the cold dark matter distribution
\begin{equation}
 C[f_h(p,a)] = - \dot{a}f_c(p)\frac{dg_h(a)}{da} = \dot{a}f_c(p)\frac{dg_c(a)}{da} .
\end{equation}
The density of the relativistic particles will be given by
\begin{equation}
 \dot{{\rho}}_h + 4\frac{\dot{a}}{a}{\rho}_h = a\dot{a}\frac{dg_c(a)}{da}{\rho}_c . \label{eq:hdm}
\end{equation}
By integrating again the perturbed distribution function, but this time taking the expansion in Legendre polynomials, a hierarchical system of equations describing the perturbations in the dark radiation temperature is obtained, where the monopole term is 
\begin{equation}
 \dot{\delta}_h =-\left(\frac{4}{3}\theta_h - \dot{a}\frac{dg_c(a)}{da}\delta_c - 4\dot{\phi}\right) ,
\end{equation}
and the dipole
\begin{equation}
 \dot{\theta}_h =\left[ k^2\left(\frac{1}{4}\delta_h - \sigma_h \right) + k^2\psi \right] ,
\end{equation}
which can be used to obtain the whole set of equations by the recurrence formula
\begin{equation}
 \dot{\mathscr{N}}_l =\frac{k}{2l+1}\left[l\mathscr{N}_{l-1} - (l+1)\mathscr{N}_{l+1} \right] , l\geq 2 ,
\end{equation}
until some order of truncation $l_{max}$, where
\begin{equation}
\mathscr{N}_{l_{max}+1}\approx\frac{\left(2l_{max}+1\right)}{k\tau}\mathscr{N}_{l_{max}}-\mathscr{N}_{l_{max}-1} . 
\end{equation}
For the desired density conversion, the \textit{ad-hoc} conversion rate must be
\begin{equation}
  \frac{dg_c(a)}{da} =\frac{\eta}{a} ,
\end{equation}
which replaced in eqs. (\ref{eq:cdm}) and (\ref{eq:hdm}), gives
\begin{equation}
 \rho_c =\frac{\rho_{c_0}}{a^3} \times e^{-\eta(a-a_i)} ,
\end{equation}
\begin{equation}
 \rho_h =\frac{\rho_{c_0}(a_i+\eta^{-1})}{a^4} \times \left[ 1 - e^{-\eta(a-a_i)} \right] ,
\end{equation}
as planned beforehand in eqs. (\ref{eq:cold}) and (\ref{eq:hot}). 
Given that the conversion is supposed to be an environmental effect related to the galaxy halo, we include a scale dependence of the form
\begin{equation}
\frac{dg_c(a,k)}{da} \rightarrow h(k,k_g) \times \frac{dg_c(a)}{da} ,
\end{equation}
\begin{equation}
\frac{dg_h(a,k)}{da} \rightarrow \left[1-h(k,k_g)\right] \times \frac{dg_h(a)}{da} ,
\end{equation}
where the term included is a smooth step function and $k_g$ is defined as the scale where the conversion reaches half of its maximum effect for each redshift. 

\section{Statistical Analysis} \label{sec:analysis}

\begin{table}
\begin{center}
\caption{Parameters of the model and their flat prior. The first six parameters in the first block belong to the so called ``vanilla'' cosmological model. The middle block contains extended standard parameters that are expected to have degeneracy with the conversion model parameters. The last block contains the parameters added by our model of dark matter conversion.}
 \begin{tabular}{l|l|p{8cm}|l|l}\hline
Type& \textbf{Parameter} & \textbf{Description}         & \textbf{Min} & \textbf{Max} \\ \hline  \hline
Vanilla & $\Omega_b h^2$    & Baryon density                    & 0.005 & 0.1  \\
&$\Omega_c h^2$    & Cold dark matter density          & 0.04  & 0.18 \\
&$\theta$          & Ratio between the sound horizon and the angular diameter distance at decoupling & 0.5   & 10   \\
&$\tau$            & Reionization optical depth & 0.01 & 0.8 \\
&$n_s$             & Spectral index of primordial power spectrum at $k=0.05h\textrm{Mpc}^{-1}$ & 0.5 & 1.5 \\
&$log[10^{10} A_s]$& Amplitude of primordial power spectrum at $k=0.05h\textrm{Mpc}^{-1}$& 2.7 & 4.0 \\ \hline
Extended&$f_\nu$           & Fraction of neutrino density related to the dark matter & 0 & 0.1 \\
&$w$               & Parameter of dark energy constant equation of state & -1.5 & 0.5 \\ \hline
Extra&$a_i$             & Scale factor when the conversion starts & 0.1 & 1.00 \\
&$\eta$            & Rate of conversion & 0 & 2 \\ \hline
 \end{tabular}\label{tab1}
\end{center}\end{table}
The statistical analysis was made by Markov Chain Monte Carlo (MCMC) using the data from the anisotropies of the cosmic microwave background measured by the WMAP, the matter power spectrum measured by the SDSS and the type Ia supernovae luminosity compiled by the Union-2. 
Best fits were obtained by conventional $\chi^2$ analysis using Bayes posterior probability description. All priors used were flat distributions over the range tested. We run the CosmoMC~\cite{Lewis:2002ah} package of MCMC to test the theoretical predictions generated by a modified version of CAMB~\cite{Lewis:1999bs}, which included the particles described by the Boltzmann equations developed in section \ref{sec:equations}. 
Besides the standard parameters for cosmology, our model add the parameters listed in the last block of table \ref{tab1}. The priors adopted are the standard for the vanilla parameters, while the prior for the parameter $\eta$ goes from vanishing conversion ($\eta=0$) to non-physical values ($\eta=2$) when all the cold dark matter is quickly converted, for the parameter $a_i$ the prior goes from conversion starting today ($a_i=1$) to non-physical values ($a_i=0.1$) when the dark radiation strongly degrades the structure formation. This choice for the extra parameters obeys the rule to stipulate a prior that encompasses a vanishing effect and non-physical results in such a way that the posterior probabilty has a well defined tail. In our case, this criteria for prior stipulation is especially designed when are included galaxy's surveys, which are our target and consequently the most affected observable. The scale conversion $k_g$ was fixed to the scale of typical galactic halo size, $k_g \equiv 0.2$ Mpc$^{-1}$. The convergence requirement followed the Gelman and Rubin $R$ statistic, which states that the variance of chains means divided by the mean of chains variances must approach of one. In this work, we reached a convergence of at least $R-1<0.05$, which could be regarded as a low convergence, but taken into account that the CosmoMC package is adjusted for statistical efficiency when fitting the standard cosmological model, we consider that for a first time fitting, the condition $R-1<0.05$ for an alternative model is acceptable.

Because there is an obvious degeneracy between neutrino density and any other kind of dark radiation, we included the fraction of dark matter in massive neutrinos $f_\nu$ among the standard parameters of $\Lambda$CDM, distributing the density of neutrinos equally among three states and fixing the number of families to $N_\textrm{eff}=3.04$. It is also expected to exist a correlation between the effects of dark matter conversion and the parameter $w$ of the dark energy equation of state ~\cite{PhysRevLett.95.221301}, for this reason this parameter was included in our analysis.

\section{Results and Discussion}\label{sec:results}

By fitting the data with theoretical predictions, we calculate chains of the likelihood for all parameter values inside the ranges delimited. The complete dataset fitted was WMAP + SDSS + SNIa and the likelihood functions adopted for each observable were the ones suggested in their respective papers. We present in table \ref{tab2} for each parameter $p$, the total mean value ($\left\langle p \right\rangle$) of its distribution and its confidence level corresponding to the $95\%$ central credible interval to represent the data.  
In table \ref{tab2}, we also present the best fit value ($\hat{p}$) of each parameter and its confidence level corresponding to the $2\sigma$ probable region. In this case, all parameters values are kept in the total best fit point. Limits not shown mean that the limit is compatible with the correspondent limit in its prior range. See ref.~(\cite{1475-7516-2007-08-021}) for different statistical approaches and their relevance in cosmology. 
In order to compare and qualify the modified model, we also present the mean and best fit points, the credible regions and confidence levels of the standard cosmological model. 
We use $\Lambda$HCDM as the acronym for the model with dark matter conversion, while keeping $\Lambda$CDM for the standard model, both extended with massive neutrinos ($m_\nu$) and an arbitrary value for the dark energy equation of state parameter ($w$). 
\begin{table}[t]
\begin{center}
\caption{Mean total values $\left\langle p \right\rangle$ with their $95\%$ central credible intervals and the best fits $\hat{p}$ with their confidence levels at $2\sigma$, for the model with conversion ($\Lambda$HCDM) and for the standard model ($\Lambda$CDM). The fits were obtained with WMAP+SDSS+SN data.}
\begin{tabular}{l | l | m{2cm} | m{2cm}  | m{2cm} | m{2cm} m{0.01cm}} \cline{1-6}
\multirow{2}{*}{Type} & \multirow{2}{*}{Parameter} &  \multicolumn{2}{c}{$\left\langle p \right\rangle$(95$\%$ CCI)}& \multicolumn{2}{|c}{$\hat{p}$ (2$\sigma$ CL)}  \\[0.8ex] \cline{3-6}
 &    &    $\Lambda$HCDM & $\Lambda$CDM & $\Lambda$HCDM & $\Lambda$CDM & \\[0.8ex] \cline{1-6} \cline{1-6}
Vanilla 
&$\Omega_b h^2$             & $0.0222_{0.0211}^{0.0234}$ & $0.0226_{0.0214}^{0.0238}$ & $0.0223_{0.0202}^{0.0245}$ & $0.0226_{0.0209}^{0.0246}$ & \\[1ex] \cline{2-6}
&$\Omega_c h^2$ 						& $0.10_{0.06}^{0.14}$       & $0.145_{0.132}^{0.160}$    & $0.07_{0.03}^{0.16}$    & $0.142_{0.122}^{0.165}$ & \\[1ex] \cline{2-6}
&$\theta$	  								& $1.035_{1.030}^{1.040}$    & $1.033_{1.028}^{1.039}$    & $1.035_{1.024}^{1.053}$    & $1.034_{1.024}^{1.042}$ & \\[1ex] \cline{2-6}
&$\tau$            					& $0.09_{0.06}^{0.12}$       & $0.08_{0.07}^{0.11}$       & $0.09_{0.04}^{0.15}$ 	     & $0.09_{0.05}^{0.13}$ & \\[1ex] \cline{2-6}
&$n_s$             					& $0.973_{0.943}^{1.002}$    & $0.99_{0.96}^{1.02}$       & $0.98_{0.92}^{1.02}$       & $1.00_{0.94}^{1.04}$ & \\[1ex] \cline{2-6}
&$log[10^{10} A_s]$					& $3.11_{3.03}^{3.18}$     & $3.13_{3.07}^{3.19}$       & $3.10_{2.98}^{3.24}$         & $3.14_{3.04}^{3.23}$ & \\[1ex] \cline{1-6}
Extended
&$f_\nu$             	      & $0.03_{}^{0.06}$           & $0.02_{}^{0.06}$           & $0.02_{}^{0.10}$              & $0.02_{}^{0.07}$ & \\[1ex] \cline{2-6}
&$-w$               				& $0.88_{1.06}^{0.74}$       & $0.83_{0.98}^{0.71}$       & $0.88_{1.31}^{0.63}$       & $0.82_{1.13}^{0.62}$ & \\[1ex] \cline{1-6}
Derived
&$\Omega_\Lambda$  	        & $0.68_{0.62}^{0.72}$       & $0.66_{0.61}^{0.71}$ 		  & $0.70_{0.54}^{0.76}$       & $0.68_{0.58}^{0.73}$ & \\[1ex] \cline{2-6}
& Age/Gyr 	  							& $13.3_{13.0}^{13.6}$       & $12.6_{12.3}^{12.8}$ 		  & $13.2_{12.8}^{13.8}$       & $12.5_{12.2}^{13.0}$ & \\[1ex] \cline{2-6}
&$\sigma_8$	  							& $0.71_{0.63}^{0.80}$       & $0.76_{0.68}^{0.85}$ 		  & $0.72_{0.58}^{0.89}$       & $0.78_{0.62}^{0.92}$ & \\[1ex] \cline{2-6}
&$z_{re}$	  								& $11_{8}^{14}$              & $11_{9}^{14}$ 					    & $11_{6}^{16}$              & $11_{7}^{15}$ & \\[1ex] \cline{2-6}
&$H_0$		  								& $69_{65}^{73}$             & $71_{67}^{75}$ 						& $71_{61}^{77}$             & $72_{65}^{77}$ & \\[1ex] \cline{2-6}
&$\sum m_\nu$(eV)  					& $0.4_{}^{0.8}$            & $0.3_{}^{0.8}$              & $0.09_{}^{1.4}$            & $0.2_{}^{1.1}$ & \\[1ex] \cline{1-6}
Extra
&$a_i$             		      & $0.6_{0.3}^{}$             & N/A                        & $0.6_{}^{}$                & N/A  & \\[1ex] \cline{2-6}
&$\eta$            					& $0.4_{}^{0.8}$ 	           & N/A                        & $0.6_{}^{1.0}$             & N/A & \\[1ex] \cline{1-6}
Extra derived
&$\Omega_h h^2$  					  & $0.06_{}^{0.10}$           & N/A                        & $0.07_{}^{0.13}$           & N/A & \\[1ex] \cline{2-6}
&$f$  					            & $0.4_{}^{0.6}$             & N/A                        & $0.5_{}^{0.7}$           & N/A & \\[1ex] \cline{1-6}
\end{tabular}\label{tab2}
\end{center}\end{table}
When comparing the best fits, the minimal $\chi^2$ for the model with dark matter conversion depreciates the fit for the following amount
\begin{equation}
 \Delta \chi^2 = \left(\chi^2_\textrm{\tiny $\Lambda$HCDM}-\chi^2_\textrm{\tiny $\Lambda$CDM}\right)/\Delta d.o.f. = 1.01/2.
\end{equation}
For two extra degrees of freedom, this number means that the best fit for $\Lambda$HCDM is worse than the $\Lambda$CDM best fit, but it is withing $1\sigma$ of confidence around the standard model best fit. The best fit point in the likelihood for the two extra parameters that characterize our model are
\begin{equation}
\eta=0.6 \ , \ \ \ a_i=0.6 \ ,
\end{equation}
where the large standard deviation prevents better precision. These values for the global best fit correspond to a rate of converted dark radiation over cold dark matter of around $50\%$ in our Universe today. 

\begin{figure}\begin{center}
\begin{center}
{\resizebox{0.9\columnwidth}{!}{\includegraphics{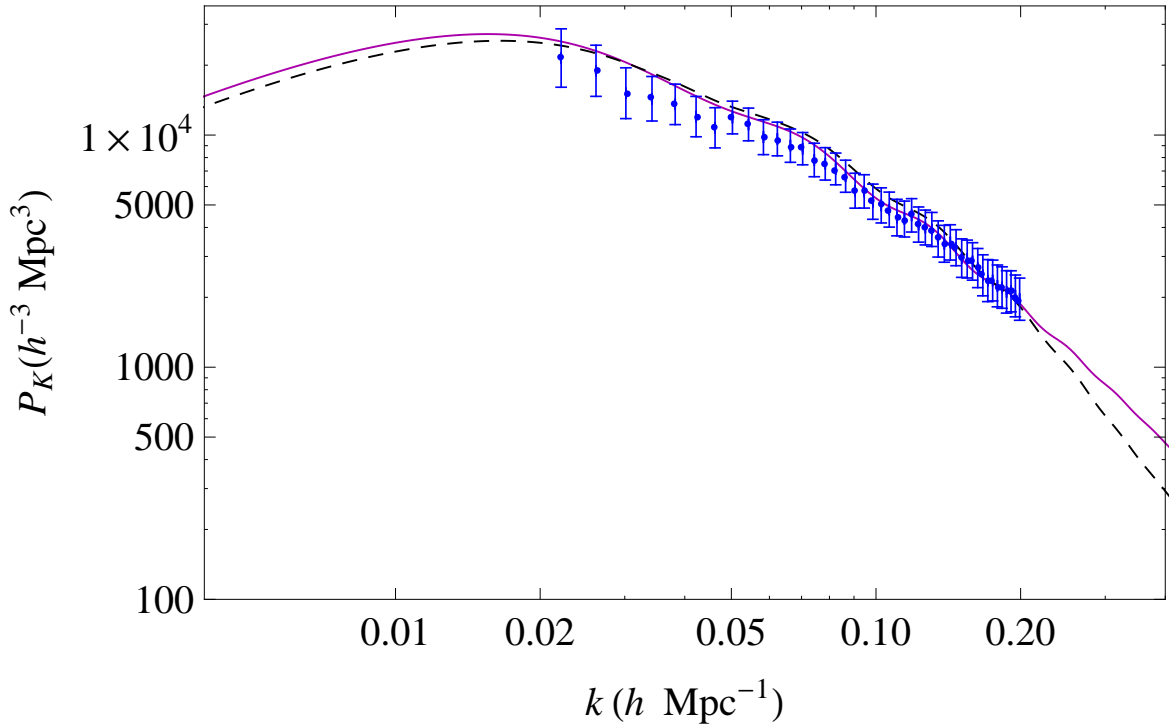}}}
 \end{center}\caption{Matter power spectra for different models containing the best fit for WMAP+SDSS+SNIa. $\Lambda$CDM (magenta continuous line) and the model with dark matter conversion (black dashed). SDSS data are shown in blue points, but only diagonal elements of the covariance matrix.}\label{fig:pk}
\end{center}\end{figure}

\begin{figure}[t]
\begin{center}
\begin{center}
{\resizebox{0.7\columnwidth}{!}{\includegraphics{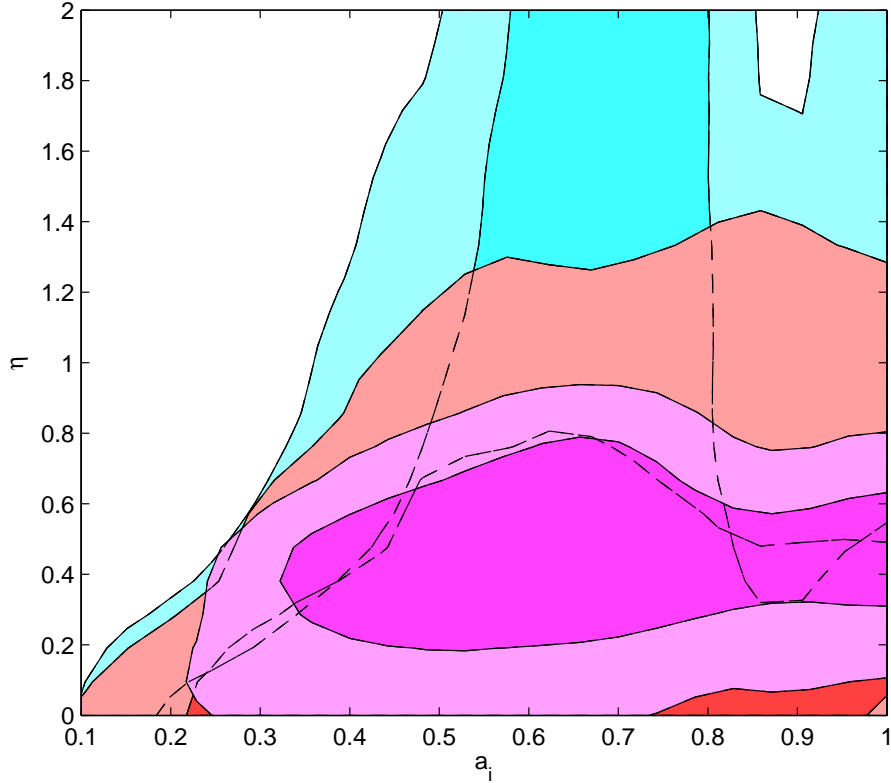}}}
 \end{center}\caption{Correlation of the parameters $a_i$ and $\eta$ at $68\%$ CL and $95\%$ CL. All others parameters are marginalized over their total mean values. The confidence regions correspond to the fits obtained with the following data sets: WMAP+SNIa (cyan), WMAP+SDSS (red) and WMAP+SDSS+SNIa (magenta).}\label{fig:marga0eta}
\end{center}\end{figure}

In figure \ref{fig:pk} we show the matter power spectrum from SDSS, the predictions for $\Lambda$CDM and the predictions for our model with dark matter conversion.  In figure \ref{fig:marga0eta} we show the correlation between the parameters $a_i$ and $\eta$.  The confidence region for the two parameters combined was obtained while fixing the value of standard parameters to their total mean. In this case the set $\{a_i,\eta\}$ excludes non-vanishing conversion $\{a_i,\eta\}\rightarrow\{1,0\}$ at $2\sigma$. However, one should not take this as a preference for conversion. The other parameters are marginalized over in their total mean, which are significantly different from the standard model mean, therefore, the conversion is needed to restore the desirable evolution of the Hubble parameter. In fact, the likelihood posterior distributions of the parameters, $a_i$ and $\eta$, are each separately compatible with vanishing conversion, even at $1\sigma$. The implication for this lack of reducibility is that the conversion can not be taken as a simple subleading effect, once it is included, the conversion must be non-vanishing to reach a best fit closer to the one obtained with the standard model.

Despite the loss of direct compatibility among the standard parameters when comparing the two models, it is interesting to note how some of them can be relaxed or tightened, such as the dark energy equation of state parameter, the cold dark matter density or the neutrino masses. Although the parallel between best fits is the main criteria to qualify an alternative model, we show  the relaxation of some parameters since it could be useful to explain some isolated effects related to specific parameters detached from the general model. 
The relaxation of neutrino masses bounds is interesting in the case one wants to add extra neutrino sterile families with masses at eV scale, such as suggested by anomalies in neutrino oscillations~\cite{Boyarsky:2009ix,PhysRevLett.107.091801}, although the allowed mass bound is still far from the scale of a few eV required by these oscillations anomalies. The conversion of cold dark matter into dark radiation decreases the dark energy equation of state parameter to lower values, to compensate for the creation of matter whose density dilutes with a higher order of the scale factor. This effect is compatible with what is expected when the neutrino masses are increased, which increases the amount of hot dark matter.

The last and most interesting side effect of our conversion model would be the smaller value for the cold dark matter density at late times, specially in the central cores where it is expected to happen the conversion. 
The decrease in density is desirable given the incompatibility between the cusp galactic centers predicted by n-body simulations and the core profile observed in all surveys~\cite{deBlok:2009sp}.
\begin{figure}[t]
\begin{center}
\begin{center}
{\resizebox{0.8\columnwidth}{!}{\includegraphics{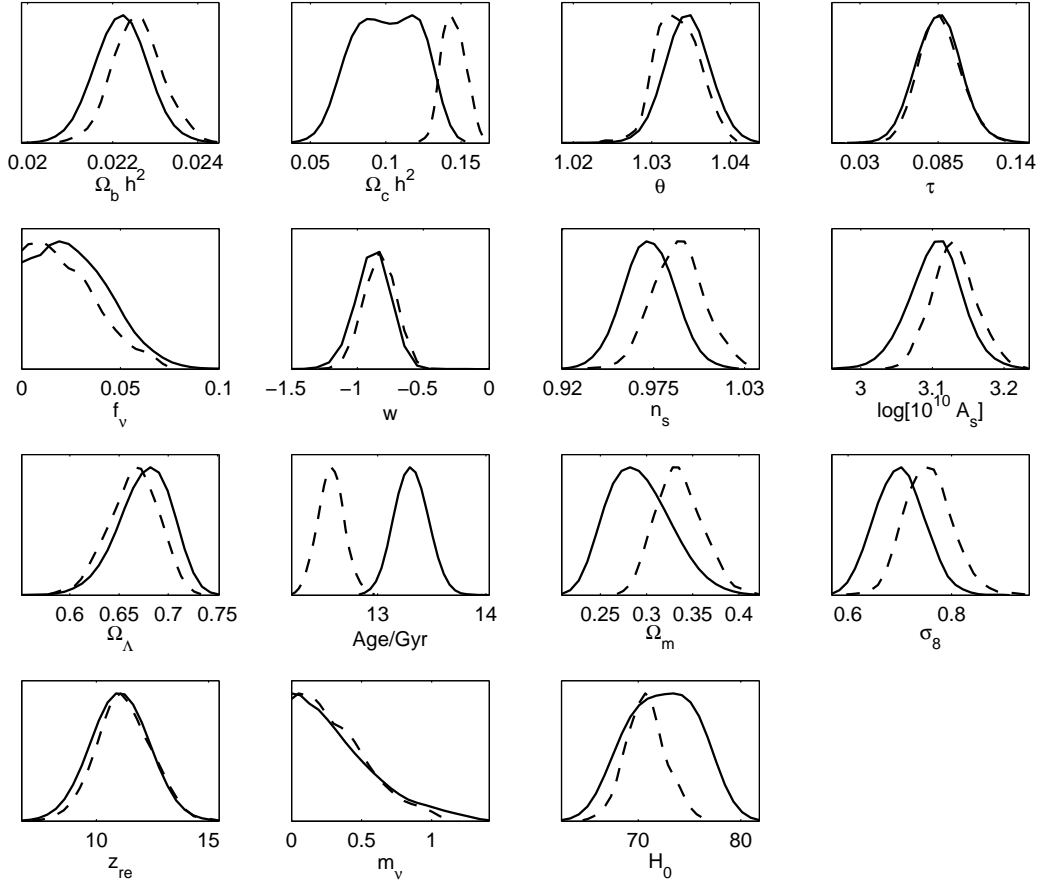}}}
 \end{center}\caption{Posterior normalized distributions for all the standard model parameters for $\Lambda$CDM (dashed line) and for the model with dark matter conversion (solid line). All others parameters are marginalized over their total mean values.}\label{fig:likelis}
\end{center}\end{figure}
For instance, in the figure \ref{fig:likelis} we show the likelihood posterior distribution for all the standard model parameters, primitive and derived ones, with dark matter conversion, and the distribution for $\Lambda$CDM. It can be seen that the largest deviation on primitive parameters happens for the total amount of dark matter density $\Omega_c$. 
Given that the conversion decreases the amount of cold dark matter at late times, it is not unexpected that the fitting procedure found a best fit with small density today for cold dark matter while keeping its early time density compatible with the found in $\Lambda$CDM scenario. Such concentrated effect in cold dark matter density is not accidental, since the conversion model was built to diminish the dark matter density at small scales without affecting the nice agreement with large scale structures.
The deviations from the other primitive parameters with respect to the scenario with no conversion are generally small and with hardly observable effects. 

\subsection{Microlensing constraints}\label{sec:microlensing}

At first look, the low counting of minihalos through microlensing \cite{Tisserand:2006zx} could be taken as an indication that the process considered in this work is highly constrained, since there would not exist enough number of dense regions to prompt the conversion. However, the cold dark matter structures considered in this work which are able to ignite the conversion are significantly bigger than in the MACHO's hypothesis ($2\pi/k_g \gg r_c \sim 5$kpc; where $r_c$ is the minihalo size in the ``S model'' of the MACHO Collaboration \cite{Alcock:2000ph}), dispensing with the requirement that such minihalos exist to initiate the conversion. Furthermore, the onset of conversion would actually weaken the constraints in the population of the minihaloes. The microlensing effect is due to the transition of the central core along the line of sight and not of the whole minihalo, thus the smoothing of the central core by decay or conversion of the cold dark matter decreases the astrometric microlensing signature of the minihaloes \cite{PhysRevD.86.043519}.

The hypothesis that minihaloes are the responsibles by the flattened rotation curve of the galaxies can be directly tested because these objects should sum up the total missing mass in each galaxy. However, it is not possible to make predictions for the population of the minihaloes taking in account the dark matter conversion into dark radiation without reconstruct all the formation history of the galaxy, which is only obtainable with dedicated n-body simulations. For this pratical impossibility, we leave the task to constraint the conversion model using microlensing for future work.

\section{Conclusions}\label{sec:conclusions}

We have tested with large scale structure data the hypothesis of an environmental conversion of cold dark matter into dark radiation that took place at late times in small scales. The effect was inspired in astrophysical processes that convert gravity binding energy in flow of relativistic particles. The main motivations to develop such phenomenological model are the particle models with composed dark matter candidates and the hints of dark radiation presence in different stages of the Universe evolution. One possible application is to address the discrepancy between the cold dark matter model and the very small scales observations, namely the core-cusp and the missing satellites issues. 

The results show that the pure cold dark matter is still a better option, although only marginally. The compatibility of the conversion with large scale structure was expected, since, aware that no conversion could take place in early times and in very large scales, we constrained the effect to happen at small scales and allowed to be triggered at late times, avoiding the most rigorous bounds in the data. In this way, the mechanism presented here works as a proposal for a model, already delimited by the available data at large scales. The main effects happen at very small scales and only observations at these scales will settle the complete portrait of the dark matter component. 
The results obtained in this work could hopefully help to direct the theoretical effort to address some issues at small scales when observational data will be available. The good agreement obtained with present data is an indication that this is not only valid, but a suitable solution.

A possible extension of this work would be to perform the n-body simulation for some galaxy clusters and put the conversion in the high density halos. Only accessing the non-linear regime of perturbation, one could explore the full capacity of such environmental mechanism of dark radiation production. Although the core-cusp and missing satellites issues are not the only observational hint for dark radiation, it is certainly the most interesting one for this model, given the factual relevance and for the possibility to be tested more directly with data.

\section*{Acknowledgments}

The authors would like to thank CNPq and CAPES for several finantial supports. DB acknowledges the computing resources from CENAPAD and CCJDR-IFGW.

\bibliographystyle{unsrt}
\bibliography{article} 

\end{document}